\documentclass[11pt,a4paper]{article}
\pdfoutput=1
\usepackage{jcappub}
\usepackage{color}
\input{colordvi.tex}
\usepackage{graphicx}
\usepackage{float}
\usepackage{wrapfig}
\usepackage{bm}
\usepackage{amsmath,amssymb}
\usepackage{subfigure}
\usepackage{verbatim}
\usepackage{makeidx}
\usepackage{mathabx}

\title{Cosmological constraints from $\mu E$ cross correlations}
\author{Atsuhisa Ota}
\affiliation{Department of Physics, Tokyo Institute of Technology,\\
Tokyo 152-8551, Japan}
\emailAdd{a.ota@th.phys.titech.ac.jp}

\abstract{
We derive general expressions of the $C^{\mu X}_l$, the cross correlation function between the cosmic microwave background spectral $\mu$ distortion and the linear perturbations in the cosmic microwave background such as the temperature perturbations and the polarizations.
The cross correlations are known as new tests for the extremely squeezed shape of primordial non-Gaussianity, which is inaccessible through the direct observations of the temperature 3-point functions. 
Our formulae are applicable to the arbitrary combinations of the scalar and the tensor perturbations, and we discuss the potential for detecting these quantities.
We provide signal-to-noise ratio of the $\mu E$ as well as $\mu T$, based on an experiment like the Primordial Inflation Explorer.
We also find the signal-to-noise ratio from the scale dependent nonlinearity.
For instance, we show that $f^{\rm loc}_{\rm NL}(k_1,k_2,k_3)=\left(k_1k_2k_3\right)^{1/3}k_0^{-1}F_0$ with $k_0=0.05{\rm Mpc}^{-1}$ can be detectable at 1$\sigma$ level even for $F_0 \sim \mathcal O(1)$.
}


\begin{document}
\maketitle

\section{Introduction}

Recent observational projects have delivered rich information on
anisotropies in the cosmic microwave background (CMB), and they have improved our understanding of the early universe~\cite{Ade:2013uln,Hinshaw:2012aka,Ade:2014xna}.
Almost scale-invariant and Gaussian primordial curvature perturbations are supported; however, we should keep in mind that we are blind, for the most part, to the primordial universe; namely, the visible scales from the CMB temperature power spectrum and bispectrum are merely the 7 e-holdings of the inflationary period. 
Therefore, it is too hasty to draw conclusions about the nature of primordial perturbations, despite the sequence of achievements. 
On the other hand, there is an essential difficulty for the further improvement 
of the observable ranges in the CMB anisotropy experiments since Silk damping erases fluctuations on angular scales smaller than $\mathcal O(1^\circ)$.
In this sense, the CMB spectral deformation induced by Silk damping can be a complementary approach to the small scale physics~\cite{Sunyaev:1970eu,Hu:1994bz,1991MNRAS.248...52B,Chluba:2012we,Khatri:2012rt,Chluba:2012gq,Khatri:2012tw,Khatri:2013dha,Clesse:2014pna,Ota:2014hha,Chluba:2014qia}.
This is because such distortions are sourced from second order effects of temperature perturbations which dissipated due to Silk damping. 
For instance, the chemical potential-type distortion to the CMB black body spectrum is estimated as $\mu\sim O(10^{-8})$ based on the almost scale-invariant and Gaussian primordial curvature perturbations~\cite{Sunyaev:1970er,Chluba:2012gq}, and the next generation of space missions is expected to detect this~\cite{Kogut:2011xw,Andre:2013afa}.
Recently, the anisotropies in the distortions were also discussed in the context of a primordial non-Gaussianity search~\cite{Pajer:2012vz,Pajer:2013oca,Ganc:2012ae,Ganc:2014wia,Ota:2014iva,Emami:2015xqa}.
Even though the chemical potentials are the thermodynamic quantities which are realized in each diffusion patch, 
they can fluctuate according to the primordial 3- and 4-point correlation functions.
This point of view was first proposed in~\cite{Pajer:2012vz}, and the authors derived the upper bound on the nonlinear parameter by calculating $C^{\mu T}_l$, the angular cross correlation between the chemical potential and the temperature perturbations.
The authors assumed a locally kinetic equilibrium system and partly included the second order effects.
We also follow their method here.

In this paper, we calculate the cross correlation functions between the $\mu$ distortions and the polarization $E$ modes as well as the temperature perturbations.
The main polarization peak is located at the scale of the last scattering surface; however this scale is invisible to the Primordial Inflation Explorer~(PIXIE) experiment whose resolutions are limited to $l\sim 84$.
Therefore, we focus on another peak by reionization at low multipoles.
It is a relatively weak signal compared to that originated at the last scattering surface; however, the line-of-sight solution of the reionized polarization $E$ modes is easily obtained~\cite{Zaldarriaga:1996ke}, and expressions for the low $l$ of $\mu$ distortions are also simple~\cite{Pajer:2013oca}.
In addition, as long as we consider the signal-to-noise ratio from reionization, the suppression factors from the reionization optical depth are less problematic since they are canceled by those in the autocorrelation functions in the denominator of the signal-to-noise ratio.
The $\mu E$ cross correlation function is useful for the following two reasons.
First, it is another observable of primordial non-Gaussianity of the squeezed triangle.
Therefore, the joint analysis with the $\mu T$ has the potential to improve the constraints.
The second is that both $\mu$ and $E$ are not affected by the late integrated Sachs-Wolfe effects due to dark energy.
It is then possible to pick up the primordial information with simple approximations.

We also comment now on the behavior of the temperature autocorrelation functions at multipole of ten thousands inspired by the small scale study of tensor perturbations.
We find that the temperature perturbations from the primordial tensor perturbations may exceed that from the scalar at extremely large $l$'s, and it is possible to investigate the primordial tensor perturbations only by observing the temperature perturbations.
Of course, the signals can glimmer excessively, and the detections may be almost hopeless due to contamination; however, such an inversion between the scalar and the tensor occurs, in principle.

We have organized this paper as follows. 
In the section \ref{section2}, we summarize the linear perturbation theory of the CMB, and we estimate the anisotropic CMB $\mu$ distortions from not only the curvature perturbations but also the gravitational waves in the section~\ref{section3}.
We write down the general expressions of the $\mu E$ and the $\mu T$ cross correlations in the section~\ref{section4} and obtained the constraints on an example of primordial non-Gaussianity in the section~\ref{section5}.
The conclusions are drawn in the final section.

\section{The linear theory of the cosmic microwave background}\label{section2}

\subsection{Stokes parameters}

We characterize the photon state by intensity and intensity contrasts called Stokes parameters.
In quantum electrodynamics, the Hamiltonian of the free electromagnetic field is written as
\begin{align}
 \hat H = \sum_{i=1,2}\int \frac{d^3p}{(2\pi)^32p}p  \hat a^\dagger_i(\mathbf p) \hat a_i(\mathbf p),
\end{align}
where $ \hat a^\dagger_i(\mathbf p)$ and $ \hat a_i(\mathbf p)$ are creation and annihilation operators of the $i$ mode polarized photons of momentum $\mathbf p$ ($p=|\mathbf p|$).
Then, the intensity operator (the energy density operator) is given as $ \hat I=  \hat H/\mathcal V$ with $\mathcal V=(2\pi)^3\delta^{(3)}(0)$ being the total volume.
On the other hand, intensity contrasts of 1- and 2-modes can be calculated by $ \hat Q= \hat P^\dagger_1  \hat I  \hat P_1- \hat P^\dagger_2  \hat I  \hat P_2$, where $ \hat P_i=\int d^3q(2\pi)^{-3}(2q)^{-1} \hat a^\dagger_i(\mathbf q)|0\rangle \langle 0| \hat a_i(\mathbf q)$ is the projection operator onto the $i$ mode polarization.
These operators determine the absolute values of the 1- and 2-modes; however, degeneracies of the polarization orientation and the phase difference remain.
Let us make a quarter turn on the plane and define a new intensity contrast operator on this frame, $\hat R^\dagger(\pi/4)  \hat Q  \hat R(\pi/4)= \hat Q'= \hat U$, where we have defined 
the $\pi/4$ rotational operator on polarization plane as $ \hat R(\pi/4)$, and this fixes the polarization direction.
The quarter wavelength plate is expressed as $ \hat P(\pi/2)=  \hat P_1+e^{-i\pi/2} \hat P_2$.
Therefore, by observing an intensity contrast behind the sequence of a quarter wavelength plate and a quarter turned polarizing plate, another intensity contrast $ \hat V= \hat P^\dagger(\pi/2)  \hat U  \hat P(\pi/2)$ can be obtained.
The operators $\hat I$, $  \hat Q$, $  \hat U$ and  $  \hat V$ are called Stokes operators (parameters), and we can identify the polarizations completely by these parameters without referring to the electromagnetic field directions on the polarization plane.
If one needs to express Stokes parameters only in the original frame, it is convenient to introduce the intensity tensor as 
\begin{align}
\hat I_{ij} = \frac{1}{\mathcal V}\int \frac{d^3p}{(2\pi)^32p}p \hat a^\dagger_i(\mathbf p)\hat a_j(\mathbf p).\label{intensitytensordef}
\end{align}
Then, we can write the operators as $ \hat I= \hat I_{11}+ \hat I_{22}$, $ \hat Q=\hat I_{11}- \hat I_{22}$, $ \hat U= \hat I_{12}+ \hat I_{21}$ and $ \hat V=i(\hat I_{12}-\hat I_{21})$.
They are also expressed as $(\hat I,\hat U,\hat V,\hat Q)=\sigma^{\mu*}_{ji} \hat I_{ij}$, where $\sigma^\mu=(1,\sigma^m)$ and $\sigma^m$ is the Pauli matrix.
Rotational dependences of the Stokes operators are manifest from (\ref{intensitytensordef}).
If one makes $\psi$ rotation on the plane $\hat I_{ij} \to {R^\dagger}_{ik}(\psi) \hat I_{kl}R_{lj}(\psi)$ with a $2\times 2$ rotational matrix $R_{ij}(\psi)$, one obtains $\hat I \to \hat I$, $\hat V \to \hat V$ and $\hat Q\pm i\hat U\to (\hat Q\pm i\hat U)e^{\mp2i\psi}$.
Then, $\hat I$ and $\hat V$ are helicity 0, and $\hat Q\pm i\hat U$ is helicity $\pm$2. 
We are interested in mixed state photon described by the density operator
\begin{align}
\hat \rho(\hat n,\mathbf x) = \int \frac{p^2 dp}{2\pi^2(2p)} \rho^{ij}(\mathbf p, \mathbf x)\hat a_j^\dagger(\mathbf p)|0\rangle\langle0|\hat a_i(\mathbf p), \label{densityop}
\end{align}
where $\hat n= \mathbf p/|\mathbf p|$, and we have assumed that the system is not dense.
$(\rho_{11}+\rho_{22})/2$ becomes an averaged Bose distribution function.
Using (\ref{densityop}), an expectation value of the intensity tensor is written as $I_{ij}={\rm Tr}[\hat \rho \hat I_{ij}]$.
If it is pure state, det$(I_{ij})=0$ so that $I,Q,U$ and $V$ are not independent; otherwise we have 4 independent parameters.
$V$ does not appear in the usual context of CMB physics since it is not generated at the last scattering surface of the Thomson scattering.

\subsection{Temperature perturbations and polarizations}

We conventionally use the dimensionless temperature perturbations and polarizations instead of the intensity: 
\begin{align}
\Theta&=\frac{I-I_0}{4I_0},\label{def:theta}\\
\Theta^{P}&=\frac{Q+iU}{4I_0},\label{def:thetap}\\
\Theta^{P*}&=\frac{Q-iU}{4I_0},\label{def:thetapstar}
\end{align}
where $I_0$ is a homogenous component of the intensity.
Let $\widetilde\Theta$, $\widetilde\Theta^{P}$ and $\widetilde\Theta^{P*}$ be the Fourier transformations of (\ref{def:theta}), (\ref{def:thetap}) and (\ref{def:thetapstar}).
Then, they originate from the primordial random variables $\xi^{(s)}_{\mathbf k}$ which are defined in the appendix~\ref{Random:val}.
To set the Fourier momentum parallel to the $z$ axis, let us transform the coordinate with the rotational matrix $R$, namely, $R\hat k=\hat k'=\hat z$ and $R\hat n=\hat n'$. 
In this frame $\hat n'$ dependence is decomposed in the following form~\cite{Zaldarriaga:1996xe}:
\begin{align}
\widetilde \Theta(\mathbf k',\hat n')=&\Theta^{S}(k,\lambda)\xi^{(0)}_{\mathbf k}+\Theta^{T}(k,\lambda)\left[(1-\lambda^2)e^{2i\phi'}\xi^{(+2)}_{\mathbf k}+(1-\lambda^2)e^{-2i\phi'}\xi^{(-2)}_{\mathbf k}\right],\label{theta_transfer1}\\
\widetilde\Theta^{P}(\mathbf k',\hat n')
=&\Theta^{SP}(k,\lambda)\xi^{(0)}_{\mathbf k}+\Theta^{TP}(k,\lambda)\left[(1-\lambda)^2e^{2i\phi'}\xi^{(+2)}_{\mathbf k}+(1+\lambda)^2e^{-2i\phi'}\xi^{(-2)}_{\mathbf k}\right],\label{thetap_transfer1}\\
\widetilde\Theta^{P*}(\mathbf k',\hat n')
=&\Theta^{SP}(k,\lambda)\xi^{(0)}_{\mathbf k}+\Theta^{TP}(k,\lambda)\left[(1+\lambda)^2e^{2i\phi'}\xi^{(+2)}_{\mathbf k}+(1-\lambda)^2e^{-2i\phi'}\xi^{(-2)}_{\mathbf k}\right],\label{thetapstar_transfer1}
\end{align}
where $\hat k \cdot \hat n=\hat k' \cdot \hat n'=\lambda$ and $\phi'$ is the rotational angle of $\hat n'$.
The linear Boltzmann equations for the transfer functions in (\ref{theta_transfer1}) and (\ref{thetap_transfer1}) are given as~\cite{Kosowsky:1994cy,Ma:1995ey}
\begin{align}
\dot \Theta^S +ik\lambda \Theta^S - \dot{\phi}+ ik\lambda \psi &=\dot \tau \left[\Theta^S-\Theta^S_0+\frac12P_2(\lambda )\Pi-\lambda v\right],\label{boltzmann1}\\
\dot \Theta^{SP}+ik\lambda \Theta^{SP} &= \dot \tau\left[\Theta^{SP} -\frac34(1-\lambda^2)\Pi\right],\label{boltzmann2}\\
\dot\Theta^T +ik\lambda \Theta^T +\dot h&=\dot \tau[\Theta^T-\Lambda]\label{boltzmann3},\\
\dot\Theta^{TP} +ik\lambda \Theta^{TP} &=\dot \tau[\Theta^{TP}+\Lambda]\label{boltzmann4},
\end{align}
where $\tau~(\dot \tau<0)$ is the optical depth, and $v$ is a velocity of the electron (baryon) fluids.
$\phi$, $\psi$ and $h$ are defined in the appendix~\ref{Random:val}.
Legendre coefficients are defined as
\begin{align}
\Theta^{X}(k,\lambda)=\sum_l(-i)^l(2l+1)P_l(\lambda)\Theta^{X}_l(k),
\end{align}
and we also introduce 
\begin{align}
\Pi &= \Theta^S_2+\Theta^{SP}_2+\Theta^{SP}_0,\\
\Lambda&=\frac3{70} \Theta^{T}_{4}+\frac17
\Theta^{T}_{2}+\frac1{10} \Theta^{T}_{0}-\frac3{70}
 \Theta^{TP}_{4}+\frac67  \Theta^{TP}_{2}-\frac35  \Theta^{TP}_{0}.
\label{eq:Lambda}
\end{align}
The integral solutions to these equations are written in the following form~\cite{Zaldarriaga:1996xe}:
\begin{align}
\Theta^X(k,\lambda,\eta_0)&=\int^{\eta_0}_0 d\eta  \mathcal S_{X}(k,\eta,\lambda)e^{-ik(\eta_0-\eta)\lambda} .\label{integral:sol}
\end{align}
where the source functions are defined as
\begin{align}
\mathcal S_{S}&=e^{-\tau }(\dot \phi+\dot\psi)+g\bigg(\Theta_0+\psi +\frac{\Pi}{4}+\frac{i\dot v}{k}+\frac{3\ddot \Pi}{4k^2}\bigg)+\ddot g\frac{3\Pi}{4k^2},\label{SS}\\
\mathcal S_{T}&=g(\dot h/\dot \tau+\Lambda),\label{ST}\\
\mathcal S_{SP}&=(1-\lambda^2)\frac34 g\Pi\label{SSP},\\
\mathcal S_{TP}&=-g\Lambda.\label{STP}
\end{align}
$g=-\dot \tau e^{-\tau}$ is a visibility function, and we have ignored $\lambda$-independent offsets which are induced by integrating by parts.

\subsection{Harmonic coefficients}

We shall classify the temperature perturbations and polarizations on the celestial sphere with angular scales; however, $\Theta^P$ and $\Theta^{P*}$ cannot be expanded by the usual spherical harmonics since they are helicities $\pm2$ on each tangent plane.
Accordingly, let us define the polarization $E$ and $B$ modes as parity odd and even parts of the polarizations with the helicity ladder operators in appendix~\ref{A}~\cite{Zaldarriaga:1996xe}:
\begin{align}
\widebar E=-\frac{\flat^2\Theta^P(\mathbf x,\hat n)+\sharp^2\Theta^{P*}(\mathbf x,\hat n)}{2},\\
\widebar B=-\frac{\flat^2\Theta^P(\mathbf x,\hat n)-\sharp^2\Theta^{P*}(\mathbf x,\hat n)}{2i}.
\end{align}
The $E$ and $B$ modes can then be expanded by the helicity 0 spherical harmonics and coefficients of $X$ are defined as
\begin{align}
a_{X,lm}=\int d\hat nY^*_{lm}(\hat n)X(\eta,\mathbf x,\hat n),\label{a_Theta}
\end{align}
where $(\eta,\mathbf x)$ is spacetime coordinate of Earth, and we usually set $\mathbf x=0$ without loss of generality.
Here, note that we conventionally define the $E$ and $B$ mode coefficients as~\cite{Zaldarriaga:1996xe}
\begin{align}
a_{E/B,lm}=\sqrt{\frac{(l-2)!}{(l+2)!}}a_{\widebar E/\widebar B,lm}.
\end{align}
In this paper we mainly focus on the cross correlations between $E$ mode and the $\mu$ distortion since the cross correlation with the $B$ modes is trivially 0 without non-standard scenarios such as parity violation.
It is more convenient to express (\ref{a_Theta}) in the form of the Fourier integral
\begin{align}
a_{X,lm}=4\pi(-i)^l\int \frac{d^3 k}{(2\pi)^3}\sum_{s}{_{-s}Y^*_{lm}}(\hat k)\xi^{(s)}_{\mathbf k}\mathcal T^{(|s|)}_{X,l}(k,\eta),\label{harmonic_coefficients}
\end{align}
where we have defined
\begin{align}
\mathcal T^{(0)}_{\Theta,l}&=\Theta^S_l\label{0:T:T}\\
\mathcal T^{(0)}_{E,l}&=\sqrt{\frac{(l+2)!}{(l-2)!}}\int^{\eta_0}_0d\eta\frac{3g\Pi}{4}\label{0:T:E}
\frac{j_l(x)}{x^2}
,\\
\mathcal T^{(2)}_{\Theta,l}&=-\sqrt{\frac{(l+2)!}{(l-2)!}}\int^{\eta_0}_0d\eta \mathcal S_T(k,\eta)\frac{j_l(x)}{x^2},\label{2:T:T}\\
\mathcal T^{(2)}_{E,l}&=-\int^{\eta_0}_0d\eta \mathcal S_{TP}(k,\eta)\mathcal {\hat E}(x)\frac{j_l(x)}{x^2},\label{2:T:E}\\
\mathcal T^{(2)}_{B,l}&=-\int^{\eta_0}_0d\eta \mathcal S_{TP}(k,\eta)\mathcal{\hat B}(x)\frac{j_l(x)}{x^2}\label{2:T:B},
\end{align}
with $x=k(\eta_0-\eta)$ and 
\begin{align}
\mathcal {\hat E}(x)&=-12+x^2-x^2\partial_x^2-8x\partial_x,\\
\mathcal {\hat B}(x)&=-2x^2\partial_x-8x.
\end{align}

\subsection{Approximate solutions to the radiative transfers}

Polarizations are mainly produced at the last scattering surface; however, these scales are invisible to  PIXIE.
Therefore, the signals originating from the reionization are more useful for our purposes.
Taking into account the reionization effects, the visibility function can be replaced by~\cite{Zaldarriaga:1996ke} 
\begin{align}
g\to (1-e^{-\tau_{\rm reio}})\delta(\eta-\eta_{\rm reio})+e^{-\tau_{\rm reio}}\delta(\eta-\eta_*), \label{replacementg}
\end{align}
where $\eta_{\rm reio}$ is the conformal time at the reionization, and $\tau_{\rm reio}=0.058$ is the reionization optical depth~\cite{Adam:2016hgk}, and we use the following set of parameters in this paper:
\begin{align}
(\eta_0,\eta_{\rm reio},\eta_*)=(1.4\times 10^4,4.5\times 10^3,2.8\times 10^2){\rm Mpc}.
\end{align}
Using (\ref{SS}) and (\ref{0:T:T}) with (\ref{replacementg}), the scalar temperature perturbations for $\eta_*\ll \eta_0$ are approximately given as
\begin{align}
\mathcal T^{(0)}_{\Theta,l}(k,\eta_0)&\simeq e^{-\tau_{\rm reio}}[\Theta^S_0(k,\eta_*)+\psi(k,\eta_*)]j_l[k(\eta_0-\eta_*)]\notag \\
&=-\frac15e^{-\tau_{\rm reio}}j_l[k(\eta_0-\eta_*)].\label{sachsfortemp}
\end{align}
Before the reionization, $\Pi(\eta_{\rm reio})\simeq\Theta^S_2(\eta_{\rm reio})\simeq \left[\Theta_0(\eta_*)+\psi(\eta_*)\right]j_2(k\eta_{\rm reio})$.
Then, (\ref{0:T:E}) yields 
\begin{align}
\mathcal T^{(0)}_{E,l}
\simeq\sqrt{\frac{(l+2)!}{(l-2)!}}\frac{-3\tau_{\rm{reio}}}{20k^2(\eta_0-\eta_{\rm{reio}})^2}j_2[k(\eta_{\rm{reio}}-\eta_*)]j_l[k(\eta_0-\eta_{\rm{reio}})],\label{reioemodescalar}
\end{align}
where we have ignored the main polarizations from the last scattering since they are negligible on large scales.
For the tensor perturbations, the approximate solutions of (\ref{boltzmann3}) are in the same forms as the scalar and are written as
\begin{align}
\Theta^T_l(k,\eta_0)
&\simeq
e^{-\tau_{\rm reio}}\frac{20\dot h(k,\eta_*)}{21\dot \tau(\eta_*)}j_l[k(\eta_0-\eta_*)]\label{appprox:sol:tens:temp}
,\\
\Lambda(k,\eta_{\rm reio})&\simeq-\frac{1}{20}\Theta^T_0(k,\eta_{\rm reio})\simeq-\frac{\dot h(k,\eta_*)}{21\dot \tau(\eta_*)}j_0[k(\eta_{\rm reio}-\eta_*)]\label{Lambda:reio},
\end{align}
where we have used (\ref{ST}).
Then, we approximately obtain the following formulae:
\begin{align}
\mathcal T^{(2)}_{\Theta,l}\simeq& -e^{-\tau_{\rm reio}}\sqrt{\frac{(l+2)!}{(l-2)!}}\frac{20\dot h(k,\eta_*)}{21\dot \tau(\eta_*)k^2(\eta_0-\eta_*)^2}j_l[k(\eta_0-\eta_*)],\\
\mathcal T^{(2)}_{E,l}\simeq&-\tau_{\rm reio}\frac{\dot h(k,\eta_*)}{21\dot \tau(\eta_*)}j_0[k(\eta_{\rm reio}-\eta_*)]\mathcal {\hat E}\frac{j_l[k(\eta_0-\eta_{\rm reio})]}{k^2(\eta_0-\eta_{\rm reio})^2},\\
\mathcal T^{(2)}_{B,l}\simeq&-\tau_{\rm reio}\frac{\dot h(k,\eta_*)}{21\dot \tau(\eta_*)}j_0[k(\eta_{\rm reio}-\eta_*)]\mathcal {\hat B}\frac{j_l[k(\eta_0-\eta_{\rm reio})]}{k^2(\eta_0-\eta_{\rm reio})^2}.
\end{align}

\section{Spectral $\mu$ distortions}\label{section3}

\subsection{Basics}

An electron plays the role of transferring photon energy so as to realize the thermodynamic state in the early universe.
In fact, it is well known that the single Compton effects are dominant just before the last scattering, and the photon number violating processes such as the pair annihilation and the double Compton scattering can be neglected after $z\sim 2\times 10^6$~\cite{Danese:1982,1991A&A...246...49B,Hu:1994bz,Chluba:2006kg}.
On the other hand, after $z\sim 5\times 10^4$, electrons are not relativistic enough and the Compton scattering in this limit (i.e. the Thomson scattering) never transfer the energy.
Therefore, the thermalization of the photon fluid when $5\times 10^4<z<2\times 10^6$ is important to the $\mu$ distortions since chemical potential is generated thanks to photon energy transfer under the number conservation.
This can be understood by the following simple thermodynamic argument~\cite{Sunyaev:1970er}.
Let us consider a photon blackbody with temperature $T_i$.
Then, the energy and the number densities are given as $\rho_i=\alpha T_i^4$ and $n_i=\beta T_i^3$ with $\alpha=\pi^2/15$ and $\beta=2\zeta(3)/\pi^2$.
Let $\rho_i\mathcal Q$ be an energy injection to this blackbody, and let us assume that the system becomes equilibrium state again after the injection, namely, $\rho_f =\rho_i(1 + \mathcal Q)$ and $n_f =n_i$.
If we respect both the energy and the number conservation laws, the new system can never be parametrized with a single temperature, that is, the new spectrum should be a Bose distribution function with a nonzero chemical potential $\mu$~\footnote{We conventionally define the dimensionless chemical potential multiplied by $-1$.}.
Let us expand the energy and the number densities in terms of $\mu$ at linear order.
Then, we can express the final state in terms of $\rho_f=\alpha T_f^4(1-A_\rho \mu)$ and $n_f=\beta T_f^3(1-A_n \mu)$, where $A_\rho=90\zeta(3)/\pi^{4}$ and $A_n=\pi^2/(6\zeta(3))$.
Solving these simultaneous equations, we have
\begin{align}
\mu=\left(\frac43 A_n-A_\rho\right)^{-1}\mathcal Q,\label{def:mudis}
\end{align}
where the numerical constant is given as $(4 A_n/3-A_\rho)^{-1}\simeq 1.40$.
On the other hand, an energy injection due to the acoustic dissipations is expressed by~\cite{Chluba:2012gq}

\begin{align}
\mathcal Q\simeq -4\int^{\eta_f}_{\eta_i} d\eta\langle \Theta \mathcal C\rangle,\label{def:Q}
\end{align}
where $\mathcal C$ is the collision term at first order and the brackets represent the ensemble average.
It was shown in~\cite{Chluba:2012gq} that $\Theta \mathcal Cp^{-2}\partial_p p^4\partial_p f^{(0)}$ is a spectral deformation without the total number change and can be regarded as an effective heating rate~\footnote{This initially has the shape of $y$ distortions~\cite{Zeldovich:1969ff}.}.
Sometimes we use the photon energy differences between $\eta_f$ and $\eta_i$ to estimate the energy injection, in other words, we calculate $\Theta\dot \Theta$ instead of $\Theta\mathcal C$, with the overdot being a partial derivative with respect to conformal time.
Actually, this can be obtained in a simple setup of photon mixing; however, this replacement cannot take into account Sachs-Wolfe effects (SW) in the cosmological setup.
For the scalar case, we can avoid this problem by setting the initial conditions for the temperature perturbations to include SW effects by hand.
On the other hand, this is crucial for tensor perturbations since initial tensor-type temperature perturbations do not exist, and photon dissipation occurs instantaneously, so that total amount of $\Theta$ is suppressed even if the energy release is occurred.
This is because the leading order of tensor-type temperature perturbations starts with the quadrupole moment and it works as a friction immediately in contrast to Silk damping of scalar perturbations.
In fact, we find that if we calculate the $\mu$ distortion with $\Theta\dot \Theta$, it is suppressed by $\dot\tau$ compared to the regular estimation given in~\cite{Ota:2014hha,Chluba:2014qia}.

The energy injection is not marginalized over the Compton mean free path (MFP) after $z\sim 5\times 10^4$.
As a result, the chemical potential may have long wavelength fluctuations if there are primordial 3- or 4-point correlations. 
Mathematically we replace $\langle \cdots \rangle $ with $\langle \cdots \rangle_{\mathbf x}$, defined as~\cite{Pajer:2012vz}
\begin{align}
\langle X \rangle_{\mathbf x} = \int d^3x' W_{r_T}(|\mathbf x'|)X(\mathbf x +\mathbf x'),\label{def:window}
\end{align}
where $W_{r_T}$ is a window function which coarse grains the neighborhood of the point $\mathbf x$.
Thus we can roughly estimate the inhomogeneous chemical potential without any complicated second order Boltzmann equations.
In a Fourier space, the window function is calculated as $\langle e^{i\mathbf k\cdot \mathbf x}\rangle_{\mathbf x}=e^{i\mathbf k\cdot \mathbf x}\widetilde W_{r_T}(|\mathbf k|)$.
Then, the inhomogeneities in the $\mu$ distortions can be written as
\begin{align}
\widetilde\mu_{\rm fr}(\mathbf k,\hat n)=-1.4\times \widetilde W_{r_T}(|\mathbf k|)\int \frac{d^3k_1d^3k_2}{(2\pi)^6}(2\pi)^3\delta^{(3)}(\mathbf k_1+\mathbf k_2-\mathbf k)\int^{\eta_f}_{\eta_i}d\eta4\widetilde{\Theta}(\mathbf k_1,\hat n)\widetilde{\mathcal C}(\mathbf k_2,\hat n).
\end{align}
After freezing out the distortions, the Boltzmann equations for the $\mu$ distortion are given as~\cite{Pajer:2013oca}
\begin{align}
\dot{\widetilde\mu} + ik(\hat k\cdot \hat n)\widetilde\mu=\dot \tau(\widetilde\mu-\widetilde\mu_0).\label{boltzmann5}
\end{align}
The monopole component of this equation becomes 
\begin{align}
\dot{\widetilde\mu}_0 - k\widetilde\mu_1=0,
\end{align}
that is, $\mu_0$ is constant at the superhorizon without sources.
The integral solutions are also obtained through the source function 
\begin{align}
\mathcal S_{\mu}=g \int \frac{d\hat n}{4\pi}\widetilde\mu_{\rm fr}(\mathbf k,\hat n).
\end{align}
The harmonic coefficient is also calculated in the same manner with that of the temperature perturbations and is given as 
\begin{align}
a_{\mu,lm}&=4\pi(-i)^l\int\frac{d^3k}{(2\pi)^3}Y^*_{lm}(\hat k)\int^{\eta_0}_0 d\eta S_\mu(\mathbf k)j_l[k(\eta_0-\eta)].\label{mualm}
\end{align}
The above discussions are applicable regardless of the origin of $\mu$ distortions.
In the following subsections we derive the harmonic coefficients for the case with the scalar and the tensor origin $\mu$ distortions.

\subsection{$\mu$ distortions from the curvature perturbations} 
The subhorizon approximate solution to (\ref{boltzmann1}) is written as~\cite{Hu:1994uz,Dodelson:2003ft}
\begin{align}
\Theta^S_1(k)&\simeq -\frac{1}{\sqrt{3}}\sin(kr_s)\exp\left(-\frac{k^2}{k_D^2}\right),\label{dodelsonsolution}
\end{align}
where we have dropped the particular solution which describes integrated Sachs-Wolfe effects.
The prefactor of (\ref{dodelsonsolution}) comes from not only the initial monopole but also the gravitational potentials which decay soon after the horizon entry. 
$k_D=(4.1/{\rm Mpc})\times 10^{-6}(1+z)^{\frac{3}{2}}$ is the comoving diffusion scale of Silk damping~\cite{Dodelson:2003ft}.  
Using $\Theta^S_2=8k\Theta^S_1/(15\dot \tau)$, we approximately obtain 
\begin{align}
\Theta^S_2(k)\simeq -\frac{8k}{15\sqrt{3}\dot \tau}\sin(kr_s)\exp\left(-\frac{k^2}{k_D^2}\right).\label{appro.:theta2}
\end{align}

The Fourier component of the collision term in the linear Boltzmann equation is written as~\cite{Kosowsky:1994cy,Ma:1995ey}
\begin{align}
\widetilde{\mathcal C}(\mathbf k_2,\hat n)&=\dot\tau\left[\Theta^S-\Theta^S_0+\frac12P_2(\hat k_2\cdot n)\Pi-(\hat k_2\cdot\hat n) v\right]\mathcal R_{\mathbf k_2}\notag \\
&\simeq -\frac{15}{4}\dot\tau P_2(\hat k_2\cdot \hat n)\Theta^S_2(k_2)\mathcal R_{\mathbf k_2},\notag \\
&\simeq \frac{2k}{\sqrt{3}}\frac{4\pi}{5}\sum_{m=-2}^2Y^*_{2m}(\hat k)Y_{2m}(\hat n)\sin(k_2r_s)\exp\left(-\frac{k^2_2}{k_D^2}\right)\mathcal R_{\mathbf k_2}
\label{collision:scalar}
\end{align}
where we have used $\Pi\simeq 5\Theta^S_2/2$ and $ v\simeq -3i\Theta^S_1$ in the tight coupling regime and omitted the higher order multipoles.
We also use (\ref{appro.:theta2}) and $P_l(\hat k\cdot \hat n)={4\pi}{(2l+1)^{-1}}\sum_{m=-l}^lY^*_{lm}(\hat k)Y_{lm}(\hat n)$ for the last line.
Combining (\ref{def:mudis}), (\ref{def:Q}) and (\ref{collision:scalar}) with $\partial_\eta k_D^{-2}=-8/(45\dot \tau)$ and averaging with respect to $\hat n$, the Fourier component of $\mu_{{\rm fr}0}$ becomes
\begin{align}
\widetilde\mu_{{\rm fr}0}(\mathbf k)\simeq &\int
\frac{d^3k_1d^3k_2}{(2\pi)^6}(2\pi)^3\delta^{(3)}(\mathbf k_1+\mathbf k_2-\mathbf k)\mathcal M^S(k_1,k_2,k)\mathcal X^S(\mathbf k_1,\mathbf k_2)
\label{delta:fourier}
\end{align}
where we have defined
\begin{align}
\mathcal M^S(k_1,k_2,k_3)=&2.8\widetilde W_{r_T}(k_3)\frac{4k_1k_2}{k_1^2+k_2^2}\notag \\
&\int^{\eta_i}_{\eta_f}d\eta\sin(k_1r_s)\sin(k_2r_s)\partial_\eta\left[\exp\left(-\frac{k_1^2+k_2^2}{k_D^2}\right)\right],\label{3.13}\\
\mathcal X^S(\mathbf k_1,\mathbf k_2)=&\frac{4\pi}{5}\sum_{m=-2}^2Y^*_{2m}(\hat k_1)Y_{2m}(\hat k_2)\mathcal R_{\mathbf k_1}\mathcal R_{\mathbf k_2}.
\end{align}
$\mathcal M^S(k_1,k_2,k_3)$ is peaky if $k_1\simeq k_2$ due to sine functions, and it is damping at $k>{r_T}^{-1}\sim k_D$ because of the window function.
Therefore, $\mathcal M^S(k_1,k_2,k_3)$ is sensitive to the configuration of $k_1\simeq k_2\gg k_3$.

\subsection{Diffusion processes for the tensor perturbations}

For the tensor perturbations, we consider the photon diffusion as a result of originally existing quadrupole anisotropies in the gravitational waves.
Here we start with the linear Einstein equations for the gravitational waves to express the anisotropic stress in terms of the tensor transfer function $h$.
The linear Einstein equation is given as
\begin{align}
\ddot h^{\rm TT}_{ij}+2\mathcal H\dot h^{\rm TT}_{ij}-\nabla^2 h^{\rm TT}_{ij}=16\pi G a^2\delta {T^i}_j^{\rm TT}.
\end{align}
The energy momentum tensor on the r.h.s. is defined as
\begin{align}
\delta {T^i}_j&=a^{-4}\int \frac{q^2dq}{2\pi^2}\int\frac{d\hat n}{4\pi}qn_in_j\delta f(\mathbf x,q,\hat n),
\end{align}
where $\delta f=-\Theta q\partial f_0/\partial q$ with $\Theta$ and $q$ being the dimensionless temperature perturbation and the comoving momentum.
See also the appendix~\ref{Random:val} for the definition of the tensor perturbations.
Let us pick up the traceless-transverse part of both sides. The equation for the transfer functions then becomes
\begin{align}
\ddot h+2\mathcal H \dot h+k^2h&=24\mathcal H^2 \pi_\gamma,\label{eq:tensortransf}
\end{align}
where we have introduced the photon anisotropic stress by
\begin{align}
\pi_\gamma &= 2\left(\frac{1}{15}\Theta^T_0+\frac{2}{21}\Theta^T_2+\frac{1}{35}\Theta^T_4\right).\label{pioftheta}
\end{align}
On the other hand, (\ref{boltzmann3}) and (\ref{boltzmann4}) yield
\begin{align}
\dot\Theta^T_l+\frac{k}{2l+1}[(l+1)\Theta^T_{l+1}-l\Theta^T_{l-1}]&=-\dot h\delta_{l0}+\dot \tau(\Theta^T_l- \Lambda\delta_{l0}),\label{multipole:temp}\\
\dot\Theta^{TP}_l+\frac{k}{2l+1}[(l+1)\Theta^{TP}_{l+1}-l\Theta^{TP}_{l-1}]&=\dot \tau(\Theta^{TP}_l+ \Lambda\delta_{l0}).\label{multipole:pol}
\end{align}
Then, $\Theta^{TP}_l\gg\Theta^{TP}_{l+1}$ and the r.h.s. of (\ref{multipole:pol}) gives $\Theta^{TP}_0\sim \Theta^T_0/4$ for $k\eta\gtrsim1$ in the tightly coupled regime, which leads to $\Lambda\sim -\Theta^T_0$/20.
Then, (\ref{pioftheta}) yields
\begin{align}
\pi_\gamma\simeq \frac{8\dot h}{63\dot \tau},\label{piofh}
\end{align}
with $\dot \tau\sim-1.6
\times 10^5{\rm Mpc}^{-1}{(\rm Mpc/\eta)}^2$ based on~\cite{Dodelson:2003ft}.
Note that we have assumed the radiation dominated epoch here and hereafter since $\mu$ era is $5\times 10^4<z<2\times 10^6$.
Then, substituting (\ref{piofh}) into (\ref{eq:tensortransf}), and defining $f= xh$ with $x=k\eta$, (\ref{eq:tensortransf}) reduces to
\begin{align}
f'' + 2\Gamma f'+\left(1-\frac{2\Gamma}{x}\right)f=0,\label{eq:modified:tensor}
\end{align}
where $\Gamma \simeq 9.0
\times 10^{-6}$ and the prime is derivative with respect to $x$.
The solutions to (\ref{eq:modified:tensor}) are expressed by the hyper-geometric functions~\cite{Chluba:2014qia}:
\begin{align}
h(x)=\frac{1}{2\sqrt{2}}e^{-x(\Gamma+\sqrt{\Gamma^2-1})}{_1F_1}\left[1+\frac{\Gamma}{\sqrt{\Gamma^2-1}},2,2x\sqrt{\Gamma^2-1}\right],
\end{align}
where we choose the solution which converges the free solution $j_0(x)$ with $\Gamma\to 0$ and have normalized $2\sqrt{2}h$ to unity at the superhorizon~\footnote{$_1F_1(\alpha,\beta;z)=\sum^\infty_{n=0}(\alpha)_n z^n/((\beta)_n n!)$ with Pochhammer symbol $(\cdots)_n$}.
Then, the transfer function for the monopole temperature perturbations and the time derivative of $h$ approximately become
\begin{align}
\Theta^T_0(k,\eta)\simeq&
2.1
\times 10^{-6}{\rm Mpc}^{-1} k\eta^2j_1(k\eta)e^{-k\eta \Gamma},\label{tens:temp:0:app}\\
\dot h(k,\eta)\simeq& 
-\frac{1}{2\sqrt{2}}kj_1(k\eta)e^{-k\eta\Gamma}
\end{align}
where we approximate $h(x)$ as $(2\sqrt{2})^{-1}j_0(x)e^{-x\Gamma}$ for simplicity.
Note that the logarithmic scale dependence is linear in contrast to the scalar case.

\subsection{$\mu$ distortion from the primordial gravitational waves}
Thus far we have obtained the transfer functions for the tensor perturbations.
Now we are ready to calculate the $\mu$ distortions originating from the primordial gravitational waves.
Using (\ref{theta_transfer1}) with (\ref{C9}) we obtain
\begin{align}
\int \frac{d\hat n}{4\pi}\widetilde\Theta(\mathbf k_1,\hat n)\widetilde{\mathcal C}(\mathbf k_2,\hat n)&\sim \Theta_0^T(k_1) \dot h(k_2) 
\mathcal X^T(\mathbf k_1,\mathbf k_2),\label{thetaC:tens}
\end{align}
where we have defined 
\begin{align}
\mathcal X^T(\mathbf k_1,\mathbf k_2)=&\frac{32\pi}{75}\sum_{m=-2}^2\left[
{_{-2}Y^*_{2m}}(\hat k_1)
{_2Y_{2m}}(\hat k_2)
(\xi^{(+2)}_{\mathbf k_1}
\xi^{(+2)}_{\mathbf k_2}
+\xi^{(+2)}_{\mathbf k_1}
\xi^{(-2)}_{\mathbf k_2})
\right.\notag \\
&\qquad\quad\quad\left.
+{_{2}Y_{2m}}(\hat k_1)
{_{-2}Y^*_{2m}}(\hat k_2)
(\xi^{(-2)}_{\mathbf k_1}
\xi^{(+2)}_{\mathbf k_2}
+\xi^{(-2)}_{\mathbf k_1}
\xi^{(-2)}_{\mathbf k_2})
\right].
\end{align}
We also have it that $\dot \Theta_0^T$ is negligible due to the suppression by $\dot \tau$.
From (\ref{def:mudis}), (\ref{def:Q}), (\ref{tens:temp:0:app}) and (\ref{thetaC:tens}) we find that 
\begin{align}
\mu_{\rm fr0}(\mathbf k)\simeq &
 \int\frac{d^3 k_1d^3 k_2}{(2\pi)^6}(2\pi)^3 \delta^{(3)}(\mathbf{k}_1+\mathbf{k}_2-\mathbf k)\mathcal M^T(k_1,k_2,k)
\mathcal X^T(\mathbf k_1,\mathbf k_2),\label{muk:tens}
\end{align}
where we have defined
\begin{align}
\mathcal M^T(k_1,k_2,k_3)=\frac{4.1\times 10^{-6}}{\rm Mpc}\mathcal W_{r_T}(k_3)\int^{\eta_f}_{\eta_i}d\eta k_1k_2\eta^2 j_1(k_1\eta)j_1(k_2\eta)e^{-(k_1+k_2)\eta\Gamma}.
\end{align}
Substituting $\eta_i\sim 0.30 {\rm Mpc}$ and $\eta_f\sim 12 {\rm Mpc}$ into (\ref{muk:tens}) and taking the ensemble average of (\ref{muk:tens}), we have a homogenous component of the $\mu$ distortion, namely, $\langle \mu(\mathbf x)\rangle \sim 1.5\times 10^{-13}$ for unity of the tensor-to-scalar ratio and the scale-invariant power spectrum~\cite{Ota:2014hha,Chluba:2014qia}.
The deviation from the results of the previous works comes from the choices of $\eta_i$ and $\eta_f$, which are roughly estimated here.

\subsection{Scalar v.s. Tensor at multipole of ten thousands}

We have some comments on the extremely small scale behavior of the temperature perturbations. 
Comparing (\ref{appro.:theta2}) with (\ref{tens:temp:0:app}), the logarithmic dependence of the tensor-type temperature perturbations are linear in $k$ in contrast to $k^2$ of the scalar one.
Therefore, the temperature perturbations from the primordial GW may naively exceed that from the curvature perturbations at extremely large multipoles even though we have no hope of detecting such weak signals.
We demonstrate a very rough estimation and we find that they cross at $l\sim \mathcal O(10^4)$, assuming that the amplitude of the low $l$ plateau is $\mathcal O(10^{-9})$, with $r=0.1$.

\section{General expressions for the $\mu E$ and $\mu T$ cross correlations}\label{section4}

Let us write the general expressions for the cross correlation functions.
(\ref{harmonic_coefficients}), (\ref{mualm}), (\ref{delta:fourier}) and (\ref{muk:tens}) immediately yield 
\begin{align}
a^A_{\mu,lm}=&4\pi(-i)^l\int \frac{d^3k_1d^3k_2d^3k_3}{(2\pi)^9}(2\pi)^3\delta^{(3)}(\mathbf k_1+\mathbf k_2-\mathbf k_3)Y^*_{lm}(\hat k_3)\mathcal X^A(\mathbf k_1 ,\mathbf k_2)\mathcal T^A_{\mu,l}(k_1,k_2,k_3),\label{4:1}
\end{align}
where the transfer functions are defined as
\begin{align}
\mathcal T^A_{\mu,l}(k_1,k_2,k_3)=\mathcal M^A(k_1,k_2,k_3)\int^{\eta_0}_0d\eta g(\eta)j_l[k_3(\eta_0-\eta)].
\end{align}
Then, combining (\ref{4:1}) with (\ref{harmonic_coefficients}), the cross correlations with the helicity $s$ component of $X$ is obtained as 
\begin{align}
C^{\mu^A X^{(s)}}_{l}=&4\pi\int \frac{d^3k_1d^3k_2d^3k_3}{(2\pi)^9} \frac{4\pi}{2l+1}\sum_{m=-l}^lY_{lm}(\hat k_3) {_{-s}Y}^*_{lm}(\hat k_3)\notag \\
& \langle \mathcal X^{A*}(\mathbf k_1,\mathbf k_2)\xi^{(s)}_{\mathbf k_3}\rangle\mathcal T_{\mu,l}^*(k_1,k_2,k_3)\mathcal T^{(|s|)}_{X,l}(k_3).
\end{align}
Assuming that the chemical potential mainly comes from curvature perturbations, the cross correlation function becomes
\begin{align}
C^{\mu^S X^{(s)}}_{l}\simeq&4\pi\int \frac{dk_+}{k_+}\frac{dk_-}{k_-}\frac{k_+^3}{2\pi^2}\frac{k_-^3}{2\pi^2} 
F^{00s}(k_-,k_-,k_+)
\notag \\&
\mathcal T_{\mu,l}^{S*}(k_-,k_-,k_+)\mathcal T^{(|s|)}_{X,l}(k_+) \left\{
\begin{array}{ll}
1,&s=0\\
-\frac{1}{2(l+2)(l-1)},& s=\pm2
\end{array}
\right\},
\end{align}
where we have used (\ref{C11}) for $s=\pm2$.
For the scalar 3-point functions, (\ref{a2}) and (\ref{powerlowtemplate}) lead to
\begin{align}
C_l^{\mu^S,X^{(0)}}\simeq&4\pi\left(-\frac{12}{5}\right)F_0\int \frac{dk_+}{k_+}\left(\frac{k_+}{k_0}\right)^{\frac{n_f}{3}}\mathcal P_{\mathcal R}(k_+)\mathcal T^{(0)}_{X,l}(k_+)\notag \\
&\int \frac{dk_-}{k_-}\left(\frac{k_-}{k_0}\right)^{\frac{2n_f}{3}}\mathcal P_{\mathcal R}(k_-)\mathcal T^{S*}_{\mu,l}(k_-,k_-,k_+)\label{result:1}.
\end{align}
The transfer functions for the $\mu$ distortion is approximately given as
\begin{align}
\mathcal T^{S*}_{\mu,l}(k_-,k_-,k_+)\simeq 2.8\left[\exp\left(-\frac{2k_-^2}{k_D^2(z)}\right)\right]^{z_i}_{z_f}\widetilde W_{r_T}(k_+) j_l[k_+(\eta_0-\eta_*)],\label{TmuSl}
\end{align}
and those for the temperature and the $E$ mode have already been given in (\ref{sachsfortemp}) and (\ref{0:T:E}).
Evaluating these numerically, we obtain~\cite{Pajer:2012vz}
\begin{align}
C_l^{\mu^S T^{(0)}}\sim 1.9\times 10^{-16} \frac{\Gamma (l-0.02)F_0}{\Gamma (l+2.02)}
\label{eg:muT}
\end{align}
where $n_f=0$.
On the other hand, the cross correlation with the $E$ mode is not like plateau.
So for example, the value of $l=10$ is given as
\begin{align}
C_{10}^{\mu^S E^{(0)}}\sim 4.4\times 10^{-21}F_0.\label{eg:muE}
\end{align}

The cross correlation function for $\mu$ distortion originated from the tensor perturbations is obtained in the same manner.
We find
\begin{align}
C^{\mu^T X^{(s)}}_{l}\simeq&4\pi\int \frac{d^3k_+d^3k_-}{(2\pi)^6}
\mathcal T_{\mu,l}^{T*}(k_-,k_-,k_+)\mathcal T^{(|s|)}_{X,l}(k_+)\notag \\
&\frac{8\pi}{15}\left[F^{-2,-2,s}(\mathbf k_-,\mathbf k_-,\mathbf k_+)+F^{-2,+2,s}(\mathbf k_-,\mathbf k_-,\mathbf k_+)\right.\notag \\
&\left.+F^{+2,-2,s}(\mathbf k_-,\mathbf k_-,\mathbf k_+)+F^{+2,+2,s}(\mathbf k_-,\mathbf k_-,\mathbf k_+)\right]\notag \\
& \left\{
\begin{array}{ll}
1,&s=0\\
-\frac{1}{2(l+2)(l-1)},& s=\pm2
\end{array}
\right\},
\end{align}
where $F^{ijk}$ is a template of the three point function of the primordial perturbations defined in (\ref{template:general}).

\section{Constraints on the nonlinear parameter}\label{section5}
Now we shall discuss the detectability of the nonlinearity.
1$\times 1$ Fisher information matrix for $F_0$ estimation is given as~\cite{Zaldarriaga:1996xe}
\begin{align}
\mathcal F= \sum_{l}\sum_{X,Y}\frac{\partial C^{\mu X}_l}{\partial F_0}{\rm Cov}^{-1}(C^{\mu X}_lC^{\mu Y}_l)\frac{\partial C^{\mu Y}_l}{\partial F_0},\label{fisher}
\end{align}
where $X,Y=T,E$ and ${\rm Cov}^{-1}$ is the inverse of the covariance matrix defined in~\cite{Zaldarriaga:1996xe}.
This gives the signal-to-noise ratio by $S/N=F_0\sqrt{\mathcal F}$.
The covariance matrix can be calculated as
\begin{align}
{\rm Cov}(C_l^{\mu X}C_l^{\mu Y})=\frac{1}{2l+1}\left[C^{\mu X}_lC^{\mu Y}_l+\left(C_l^{\mu\mu}+C_l^{\mu\mu,N}\right)\left(C_l^{XY}+C_l^{XY,N}\right)\right],\label{CovXX}
\end{align}
where $N$ represents a noise.
The instrumental noises of the $EE$ and the $TT$ autocorrelation functions are negligible compared to the signals; however, the noise is dominant for the $\mu\mu$ autocorrelation.
For experiments like PIXIE~\cite{Kogut:2011xw}, the noise can be parametrized by
\begin{align}
C^{\mu\mu,N}_l\simeq w^{-1}_{\mu}e^{l^2/l^2_{\rm max}},
\end{align}
where $w^{-1/2}_{\mu}\simeq\sqrt{4\pi}\times 10^{-8}$, $l_{\rm max}\simeq 84$ and the upper bound of the sum in (\ref{fisher}) is set to $2l_{\rm max}$ for simplicity.
Under these conditions, $F_0<\mathcal O(10^4)$ implies that the first term on (\ref{CovXX}) is also negligible.
Ignoring the polarizations, we obtain the following signal-to-noise ratio:
\begin{align}
\left(\frac{S}{N}\right)_{T}\simeq 5.4 \times 10^{-4} \left(\frac{\sqrt{4\pi}\times 10^{-8}}{w_\mu^{-1/2}}\right)bF_0,\label{constmuT}
\end{align}
where $b$ represents the scale dependence, and $b=1$ for the scale-invariant nonlinear parameter~($n_f=0$).
Although this upper bound looks very weak compared to what is already known, the observing scale is far smaller than in previous cases, and this is an independent constraint.
Our estimation is a bit smaller compared to the original result in~\cite{Pajer:2012vz}.
This is because we take the different normalization conditions which are defined in~\cite{Ade:2013uln}, and the spectral index is not set at unity but at $0.96$; however, even when excluding such modifications, we find a small discrepancy with their results.
This is because we do not take into account the neutrino anisotropic stress~\cite{Chluba:2013dna}.
We find that it is in broad agreement with the results of the previous work if we replace the transfer functions to include the effect~\footnote{Drawing from~\cite{Chluba:2013dna}, we multiply 0.81 to the $\mu$ distortion to take into account this effect.}.
$b$ is enhanced if the bispectrum grows at the small scales.
Assuming the power law type of local-type non-Gaussianity given in (\ref{a2}), we find that $b=\{2.1\times 10^2,1.4\times 10^5\}$ for $n_f=\{1.0,2.0\}$.
Therefore, the scale dependent local-type non-Gaussianity with $F_0\sim 8$ and $n_f\simeq 1$ is detectable at 1$\sigma$ level.
We also evaluate the signal-to-noise ratio from the $\mu E$ alone and the magnitude is given as
\begin{align}
\left(\frac{S}{N}\right)_{E}\simeq 2.4 \times 10^{-4} \left(\frac{\sqrt{4\pi}\times 10^{-8}}{w_\mu^{-1/2}}\right)F_0,\label{constmuT}
\end{align}
which is less than half of that from the $\mu T$ alone.
Here we take into account only the contributions from the reionization since the noise on the $\mu\mu$ autocorrelation suppresses the signal-to-noise ratio above $l_{\rm max}$; however, we should keep in mind that those contributions coming from recombination can be comparable around $l\sim 60$.
In this context, the main polarizations should also be included to be more precise if the resolution (i.e. $l_{\rm max}$) becomes even better.
Next, let us combine the polarizations and the temperature perturbations.
Based on the same assumptions, we approximately obtain
\begin{align}
{\rm Cov}(C_l^{\mu T}C_l^{\mu E})\simeq \frac{1}{2l+1}C_l^{\mu\mu,N }C_l^{TE}.\label{CovXY}
\end{align}
Using this with (\ref{fisher}) and (\ref{CovXX}), one finds that the total signal-to-noise ratio is not improved from the case with $\mu T$ alone.
This is apparent from the following relation:
\begin{align}
\frac{C_l^{\mu E}}{C_l^{\mu T}}=\frac{C_l^{T E}}{C_l^{T T}}.\label{consistency}
\end{align}
This equation is hold as long as we assume $n_f=0$ and Sachs-Wolfe approximation since the form of the spherical Bessel function in (\ref{TmuSl}) is the same with that of the temperature transfer functions.
The $\mu E$ is not  independent quantity from the $\mu T$ if we know the $TE$ and the $TT$.
In other words, we can think of (\ref{consistency}) as a consistency relation for the $\mu X$ cross correations.
In our analysis, we consider the low $l$ region, and we only take into account the polarizations coming from the reionization; however, the consistency relation is expected to be approximately hold even for the small $l$, otherwise the signal-to-noise ratio can be improved.

Finally, we shall comment on the $S/N$ ratio of the ideal experiment.
As was discussed in~\cite{Pajer:2012vz}, in a limit of $w_\mu \to \infty $, the authors replace $C_l^{\mu\mu,N}$ with $C_l^{\mu \mu}\sim \mathcal O(10^{-29})$ and find that the sensitivity to $f_{\rm NL}$ reaches $\mathcal O(10^{-3})$ in principle.
On the other hand, we do have additional comment on this matter.
$C_l^{\mu \mu}\sim \mathcal O(10^{-29})$ is actually, the contribution from the disconnected 4-point function; however, the existing $f_{\rm NL}$ leads to a nonzero $\tau_{\rm NL}$, according to the Suyama-Yamaguchi inequality~\cite{Suyama:2007bg}.
This means that for $f_{\rm NL}>10^{-3}$, $C^{\mu\mu}_{l}\sim 10^{-23}\tau_{\rm NL}/l(l+1)$ is dominant.
Then, the upper bound of the signal-to-noise ratio is $f_{\rm NL}$ independent for the equality case~(i.e. the single field case) since it cancels with that in a numerator.
We roughly estimated $S/N\sim \mathcal O(100)$, ignoring the high $l$ region.
This means that the signal-to-noise ratio can be significant and robust up to $f_{\rm NL}>\mathcal O(10^{-3})$ in the cosmic variance limit if we ignore secondary effects.

\section{Conclusions}\label{section6}
	
We derived general expressions of the cross correlation functions between the $\mu$ distortions and the linear perturbations of the CMB such as the temperature perturbations and the polarization $E$ mode.
Our templates are applicable to arbitrary combinations of the tensor and the scalar perturbations.
The cross correlation functions are useful for primordial non-Gaussian search, and we obtained the signal-to-noise ratio for an experiment like PIXIE.
The results are in broad agreement with~\cite{Pajer:2012vz} for the scale-invariant $f^{\rm loc}_{\rm NL}$.
Then, we investigated several possibilities for enhancing the signals.
One approach is to assume that the non-Gaussianity is highly squeezed and grows on small scales.
We evaluated a case with the power law type of scale dependent non-Gaussianity, and found it detectable at 1$\sigma$ level with the unity of $n_f$ and $F_0\sim 8$.
On the other hand, we tried another approach, namely, the joint analysis of $\mu E$ and $\mu T$.
Combining these analysis, we found that the signal-to-noise ratio is not improved compared to that of the $\mu T$ alone.
Instead, we found a consistency relation for the cross correlation functions with the $\mu$ distortion.
We only focused on low $l$ behavior and have used the approximate solutions which are not applicable to small $k$'s.
It is sufficient at this stage since the resolutions are up to $l\simeq 84$ even for the latest technology; however, a wider range of analysis will be required for further understanding in the future.
This problem will be discussed elsewhere~\cite{ota}.

\acknowledgments 

We would like to thank Jens Chluba and Masahide Yamaguchi for reading our manuscript and for their many helpful comments.
We also would like to thank Enrico Pajer, Hiroyuki Tashiro and Atsushi Naruko for their helpful comments.
We would like to thank Eiichiro Komatsu for useful discussions.
An anonymous referee also gave us important comments.
The author is supported by a Grant-in-Aid for JSPS Fellows.

\appendix

\section{Ramdom variables}\label{Random:val}

The statistics of the primordial perturbations are characterized by random variables $\xi^{s}_{\mathbf k}$, where $\mathbf k$ is the Fourier momentum and $s$ is the helicity defined around $\mathbf k$.
Let $\xi^{(0)}_{\mathbf k}$ be equal to $\mathcal R_{\mathbf k}$, the gauge invariant curvature perturbation on the comoving slice.
Then, the scalar perturbations in the conformal Newtonian gauge are given as
\begin{align}
g_{00}&=-a^2(1+2\psi),\\
g_{0i}&=g_{i0}=0,\\
g_{ij}&=a^2(1-2\phi)\delta_{ij},
\end{align}
where Fourier components are related at the superhorizon as follows:
\begin{align}
\phi_{\mathbf k}=\psi_{\mathbf k}=-\frac{3+3\omega}{5+3\omega}\xi^{(0)}_{\mathbf k},
\end{align}
 where $\omega$ is the fraction of the pressure and the energy density, and we have ignored the neutrino anisotropic stress for simplicity.
On the other hand, let us introduce the tensor perturbations by 
\begin{align}
g_{00}&=-a^2,\\
g_{0i}&=g_{i0}=0,\\
g_{ij}&=a^2(\delta_{ij}+h^{\rm TT}_{ij}),
\end{align}
where $h^{\rm TT}_{ij}$ satisfies traceless-transverse condition, that is, $\partial_i h^{\rm TT}_{ij}=h^{\rm TT}_{ii}=0$.
Then, the gravitational waves are decomposed into
\begin{align}
h_{ij,\mathbf k}=2\sqrt{2}h\sum_{s=\pm 2}\xi^{(s)}_{\mathbf k}e^{(s)}_{ij}(\mathbf k),
\end{align}
where $\pm$ represents the helicity $\pm 2$, $e^{(s)}_{ij}$ is the gravitational wave polarization basis whose normalization conditions are given as $e^{(+)}_{ij}e^{(-)}_{ji}=2$ and $e^{(+)}_{ij}e^{(+)}_{ji}=e^{(-)}_{ij}e^{(-)}_{ji}=0$, and $2\sqrt{2}h$ is set to unity at the superhorizon~\footnote{$2\sqrt{2}$ is necessary to make it consistent with $h$ in (\ref{boltzmann3}).}.
The template of the initial power spectrum can be written as
\begin{align}
\langle \xi^{s_1}_{\mathbf k_1}\xi^{s_2}_{\mathbf k_2}\rangle&=(2\pi)^3\delta^{(3)}(\mathbf k_1+\mathbf k_2)P^{s_1s_2}(k_1),
\end{align}
where we have introduced
\begin{align}
P^{00}(k)&=P_{\mathcal R}(k)=\frac{2\pi^2}{k^3}\mathcal P_{\mathcal R}(k),\\
P^{\pm2,\mp2}(k)&=\frac{P_{T}(k)}{4}=\frac{2\pi^2}{k^3}\cdot \frac{\mathcal P_{T}(k)}{4}.
\end{align}
We also define the spectral indices by $n_s-1=d\ln \mathcal P_{\mathcal R}/d\ln k$, $n_T=d\ln \mathcal P_{T}/d\ln k$, and $\mathcal P_T(k_0)=r\mathcal P_{\mathcal R}(k_0)$ with $r$ being the tensor-to-scalar ratio at the pivot scale.
In this paper we use $k_0=0.05{\rm Mpc}^{-1}$, $\mathcal P_{\mathcal R}(k_0)=2.2\times 10^{-9}$ and $n_s=0.96$.
Local type non-Gaussianity in $\xi^s_{\mathbf k}$'s can be also parametrized as 
\begin{align}
\langle \xi^{s_1}_{\mathbf k_1}\xi^{s_2}_{\mathbf k_2}\xi^{s_3}_{\mathbf k_3}\rangle&=(2\pi)^3\delta^{(3)}(\mathbf k_1+\mathbf k_2+\mathbf k_3)F^{s_1s_2s_3}(\mathbf k_1,\mathbf k_2,\mathbf k_3).\label{template:general}
\end{align}
For the scalar-scalar-scalar with the rotational invariance, it is written as
\begin{align}
F^{000}(k_1,k_2,k_3)=-\frac65f^{\rm loc}_{\rm NL}(k_1,k_2,k_3)\left[P_{\mathcal R}(k_1)P_{\mathcal R}(k_2) + 2 {\rm perms.}\right].\label{a2}
\end{align}
The power law type of the scale dependent non-Gaussianity can be parametrized as~\cite{Sefusatti:2009xu}
\begin{align}
f^{\rm loc}_{\rm NL}(k_1,k_2,k_3)=\left(\frac{k_1k_2k_3}{k^3_0}\right)^{\frac{n_f}{3}}F_0,\label{powerlowtemplate}
\end{align}
where $F_0\lesssim \mathcal O(1)$ is the local-type nonlinear parameter which is already constrained in CMB anisotropy scales~\cite{Ade:2013uln}.

\section{Helicity ladder operators}\label{A}

We usually classify the polarizations according to the ``helicity'' determined by the rotational dependence around the photon momentum.
Let $\{e^\mu_1\partial_\mu,e^\mu_2\partial_\mu\}$ be an orthonormal basis on a two dimensional tangent vector space.
A rotation of an angle $\psi$ is then the linear transformation given as
\begin{align}
\left(\begin{array}{c}e'^\mu_1 \\e'^\mu_2\end{array}\right)=\left(\begin{array}{cc}\cos\psi & \sin\psi \\-\sin\psi & \cos\psi\end{array}\right)\left(\begin{array}{c}e^\mu_1 \\e^\mu_2\end{array}\right).\label{coordinate:transf:tang}
\end{align}
Here $\eta$ is said to be helicity $s$ if it transforms as $\eta \to \eta'=\eta e^{-is\psi}$ corresponding to the transformation above.
For example, $m^\mu=e^{i\gamma}(e^\mu_1+ie^\mu_2)$ with $\gamma$ being a phase factor is helicity 1 and polarizations on the tangent plane on a celestial sphere are helicity $\pm2$.
One may consider that the complex tangent vector $m^\mu\partial_\mu$ is a helicity raising operator since the helicity is 1 by itself; however, we need to introduce a connection because $\psi$ may have spacial dependences.
A similar prescription to the local gauge theory is applicable, and invariant expressions for raising operator and lowering one on helicity $s$ quantities are~\cite{Dray:1984gy}
\begin{align}
\sharp &= m^\mu\partial_\mu-\frac s2 m^{*\mu} m^\nu \nabla_\nu m_\mu,\\
\flat &= m^{*\mu}\partial_\mu+\frac s2 m^\mu m^{*\nu} \nabla_\nu m^*_\mu,
\end{align}
where $\nabla_\nu$ is the covariant derivative on the plane.
Practically, we are interested in the tangent space on $S^2$.
In this case the metric tensor is given as $g_{\mu\nu}dx^\mu dx^\nu=d\theta^2 + \sin^2\theta d\phi^2$ so that we obtain the following coordinate expressions of $\sharp $ and $\flat$~\cite{Dray:1984gy}:
\begin{align}
\sharp &= e^{i\gamma}\left[\partial_\theta +\frac{i}{\sin\theta}\partial_\phi -s\left(\cot\theta+i\partial_\theta \gamma -\frac{\partial_\phi \gamma}{\sin\theta}\right)\right],\\
\flat &=e^{-i\gamma}\left[\partial_\theta -\frac{i}{\sin\theta}\partial_\phi +s\left(\cot\theta-i\partial_\theta \gamma -\frac{\partial_\phi \gamma}{\sin\theta}\right)\right].
\end{align}
If we take $\gamma=\pi$, these are simplified to~\cite{Newman:1966ub}
\begin{align}
\sharp &=-\sin^s\theta\left(\partial_\theta +\frac{i}{\sin\theta}\partial_\phi\right)\sin^{-s}\theta,\label{paigesharp}\\
\flat &=-\sin^{-s}\theta\left(\partial_\theta -\frac{i}{\sin\theta}\partial_\phi\right)\sin^{s}\theta\label{paigeflat}.
\end{align}
Imposing $\gamma=\pi-\phi$, the expressions on the stereographic coordinate $\zeta=e^{i\phi}\cot\frac\theta2$ are also simplified to $\sharp = 2P^{1-s}\partial_\zeta P^s$ and $\flat = 2P^{1+s}\partial_{\zeta^*} P^{-s}$ with $P=(1+\zeta\zeta^*)/2$~\cite{Newman:1966ub}.
This form of ladder operators is useful for showing the completeness, and note that the phase factor is different from (\ref{paigesharp}) and (\ref{paigeflat}).
It is also useful to detail the operations on helicity $\pm 2$ quantities in terms of $\lambda=\cos\theta$:
\begin{align}
\sharp^2&=\left(\partial_\lambda-\frac{i\partial_\phi}{1-\lambda^2}\right)^2(1-\lambda^2),\label{sharp:onminus2}\\
\flat^2&=\left(\partial_\lambda+\frac{i\partial_\phi}{1-\lambda^2}\right)^2(1-\lambda^2).\label{flat:onplus2}
\end{align}

\section{$D$ matrix and spin-Spherical harmonics}\label{AC}

The $D$ matrix and spin-spherical harmonics are useful when classifying the tensor perturbations.
Here we summarize convenient formulae in our calculations.
A rotational $D$ operator in the fixed frame is defined as~\cite{Dray:1986}
\begin{align}
\hat D(\alpha,\beta,\gamma)= \hat R_z(\alpha)\hat R_y(\beta)\hat R_z(\gamma),
\end{align}
where $\hat R_i(\alpha)$ is a rotational operator of the angle $\alpha$ around the $i$ axis.
Then, a matrix representation is defined with angular momentum eigenstates:
\begin{align}
D^{(l)}_{mm'}(\alpha,\beta,\gamma)=\langle l,m|\hat D(\alpha,\beta,\gamma)|l,m'\rangle. 
\end{align}
This is an unitary matrix, namely
\begin{align}
D^{(l)}_{mm'}(\alpha,\beta,\gamma)=D^{(l)*}_{m'm}(-\gamma,-\beta,-\alpha).
\end{align}
Using the $D$ matrix, a spin-spherical harmonics is defined as
\begin{align}
{_sY_{lm}}(\theta,\phi)=\sqrt{\frac{2l+1}{4\pi}}D^{(l)*}_{m,-s}(\phi,\theta,0)=\sqrt{\frac{2l+1}{4\pi}}D^{(l)*}_{-sm}(0,\theta,\phi).
\end{align}
A rotation of ${_sY_{lm}}$ is expressed by
\begin{align}
{_sY_{lm}}(R \hat n)=\sum_{m'}{_sY_{lm'}}(\hat n)D^{(l)}_{m'm}(-\gamma,-\beta,-\alpha)=\sum_{m'}D^{(l)*}_{mm'}(\alpha,\beta,\gamma){_sY_{lm'}}(\hat n),
\end{align}
where $R$ is the $3\times3$ matrix representation of $D(\alpha,\beta,\gamma)$.
Now let us consider
\begin{align}
\hat k&=(\sin\theta\cos\phi,\sin\theta\sin\phi,\cos\theta),\\
\hat z&=(0,0,1),
\end{align}
with $(\alpha,\beta,\gamma)=(0,-\theta,-\phi)$.
Then, we have $\hat k\to R\hat k=\hat z$ and 
\begin{align}
{_sY_{lm}(R\hat n)}=(-1)^{m}\sqrt{\frac{4\pi}{2l+1}}\sum_{m'=-l}^l{_{-m}Y^*_{lm'}}(\hat k){_{s}Y_{lm'}}(\hat n).
\end{align}
This yields 
\begin{align}
(1-\lambda^2)e^{2i\phi}&=\sqrt{\frac{32\pi}{15}}Y_{22}(R\hat n)\notag \\
&=\frac{4\pi}{5}\sqrt{\frac{8}{3}}\sum_{m=-2}^2{_{-2}Y^{*}_{2m}(\hat k)Y_{2m}(\hat n}),\label{C9}\\
(1-\lambda)^2e^{2i\phi}&=8\sqrt{\frac{\pi}{5}}{_2Y_{22}}(R\hat n)\notag\\ 
&=\frac{16\pi}{5}\sum_{m=-2}^2{_{-2}Y^{*}_{2m}(\hat k){_2Y_{2m}}(\hat n}),\label{C10}
\end{align}
and
\begin{align}
\sum_{m=-l}^l{_{\pm 2}Y}^*_{lm}(\hat k)Y_{lm}(\hat k)
=-\frac{2l+1}{4\pi} \frac{1}{2(l+2)(l-1)}.\label{C11}
\end{align}

\end{document}